\documentclass[%
 reprint,
 superscriptaddress,
 amsmath,amssymb,
 aps,
]{revtex4-2}

\usepackage{graphicx}
\usepackage{dcolumn}
\usepackage{bm}

\usepackage{color}
\newcommand{\TJKK}[1]{#1}

\begin{document}

\preprint{APS/123-QED}

\title{\TJKK{Understanding how T helper cells learn to coordinate effective immune responses through the lens of reinforcement learning}}

\author{Takuya Kato\normalfont\textsuperscript{$\ddagger$,}}
  \email{takuya.kato.origami@gmail.com}
 \affiliation{
 Department of Mathematical informatics, Graduate school of Information and Science,\\
 The University of Tokyo, 7-3-1, Hongo, Bunkyo-ku, 113-8654, Japan}
\thanks{TK and TJK designed the research, conducted mathematical analysis, and wrote the paper. TK also conducted numerical simulations.}
\author{Tetsuya J. Kobayashi\normalfont\textsuperscript{$\ddagger$,}}
  \email{tetsuya@mail.crmind.net}
 \affiliation{
 Institute of Industrial Science, The University of Tokyo,\\
 4-6-1 Komaba, Meguro-ku 153-8505, Tokyo, Japan}
 \affiliation{
 Department of Mathematical informatics, Graduate school of Information and Science,\\
 The University of Tokyo, 7-3-1, Hongo, Bunkyo-ku, 113-8654, Japan}
 \affiliation{
 PRESTO, Japan Science and Technology Agency (JST),\\
 4-1-8 Honcho Kawaguchi, Saitama 332-0012, Japan}
 \affiliation{
 Universal Biology Institute, The University of Tokyo,\\
 7-3-1, Hongo, Bunkyo-ku, 113-8654, Japan}

\date{\today}

\begin{abstract}
The adaptive immune system \TJKK{of vertebrates} can detect, respond to, and \TJKK{memorize} diverse pathogens from past experience. While the clonal selection of T helper (Th) cells is the simple and established mechanism to better recognize new pathogens, the question that still remains unexplored is how the Th cells can \TJKK{acquire} better ways to bias \TJKK{the responses} of immune cells for eliminating pathogens more efficiently by translating the recognized antigen information \TJKK{into regulatory signals}. 
In this work, we address this problem by associating the adaptive immune network organized by the Th cells with reinforcement learning (RL).
By employing recent advancements of \TJKK{network-based} RL, we show that the Th immune network can acquire the association between antigen patterns of and the effective responses to pathogens.
Moreover, the clonal selection as well as other inter-cellular interactions are derived as a learning rule of the network. 
We also demonstrate that the stationary clone-size distribution after learning shares characteristic features with those observed experimentally. 
Our theoretical framework may contribute to revising and renewing our understanding of adaptive immunity as a learning system.
\end{abstract}

\maketitle


\section{Introduction}

The adaptive immunity \TJKK{of vertebrates} is a complex \TJKK{adaptive} system. 
The system constantly \TJKK{adapts} to intruding pathogens by orchestrating the populations \TJKK{and responses of} diverse immune cells, each type of which can have distinct roles \cite{abbas2014cellular,murphy2016janeway,annunziato20153}. 
For example, effector cells (innate cells, T killer cells, a part of innate lymphoid cells, B cells, etc.) are responsible for executing intrinsic pathogen-specific responses, whereas T helper (Th) cells mainly control and bias the activities of these effector cells. 
\TJKK{The diversity and activity of immune cells are modulated over the organisms' lifetimes} through intercellular communications via hundreds of chemical messengers and subsequent adaptive changes in the population sizes or phenotypic states \cite{Satija2014TrendsImmunol,Villani2018AnnuRevImmunol,Kveler2018NatBiotech, brodin2015variation}.
\TJKK{Even though young children are susceptible to infections \cite{dowling2014ontogeny, simon2015evolution, von2010farm}, they may develop higher resistance to infections through the modulation.
As evidenced by the vaccination and immunization \cite{baumgartner2012tcr, rattan2019protein, tian2016changes}, such modulation may be achieved as an adaptive response to previous infections, which can be regarded as a type of learning from experience.}

Despite the availability of latest experimental technologies, revealing the principles of such complex learning dynamics is still intricate because the immunological \TJKK{dynamics} is shaped and organized by the collective interactions of the entire immune cell population, which prevents us from simply reducing the problem down to the mere existence of specific cell types or molecules. In order to comprehend a complex learning system in neuroscience, David Marr highlighted the importance of characterizing the system at three levels \cite{marr2010vision, hassabis2017neuroscience}: the goal of the system (the computational level), the process and computation to realize the goal (the algorithmic level), and the physical implementation of the process (the implementation level).

In the past decades, a substantial amount of effort has been devoted to understand the immune system, especially at the implementation level \cite{abbas2014cellular,murphy2016janeway}. 
Cellular and molecular immunology has identified hundreds of phenotypically and functionally distinct immune cells and associated molecular markers\cite{Villani2018AnnuRevImmunol}.
Concurrently, tens of cytokines and chemokines have been discovered as chemical messages to coordinate the communications between immune cells \cite{rieckmann2017social,Kveler2018NatBiotech}. 
Moreover, with the advancement of high-throughput sequencing, it is now possible to measure the diversity of T and B cells, which constitutes an integral part of immunological recognition and memory \cite{heather2017high, ruggiero2015high}. 
Despite the accumulation of such knowledge  at the implementation level, our understanding of the immune system at the algorithmic and computational levels lags far behind and still remains limited to conceptual theories such as the clonal selection theory \cite{burnet1976modification,Perelson1989Immunolrev}.
In the face of the revealed complexity, the theory is neither sufficiently descriptive nor quantitative to draw new insights\cite{Bell1970Nature} and should be renewed to have a greater explanatory and predictive power by being endowed with a firm mathematical basis\cite{DeBoer1997IntImmunol,Mayer2015PNAS,Mayer2018arXiv,Mayer2019PNAS}.
In particular, most of theoretical works still focus only on antigen-induced T cell selection even though activated innate immune cells convey information to T cells about the origin and nature of antigens and pathogens via co-stimulation and cytokine signals\cite{Jain2017JImmunol}.
The problem that still remains unsolved is how Th cells are \TJKK{modulated} by these different signals not only to recognize antigens but also to induce and bias the activities of the groups of effector cells for evicting pathogens more efficiently by translating the recognized antigen information.

To this end, we revised the concept of immunological learning and bestowed it with a modern mathematical basis by focusing on the computational and algorithmic levels. 
At the computational level, the goal of the system may be to learn better ways from past experiences to bias the activities of the effector cells in response to infections, so as to evict the infected pathogens more promptly and specifically. 
We formulated this process as a reinforcement learning (RL) problem described using a Markov decision process (MDP) \cite{Neftci2019NatMachInt,sutton1998reinforcement,Puterman2014Book}.

\begin{figure}[tbhp]
\centering
\includegraphics[width=0.98\linewidth]{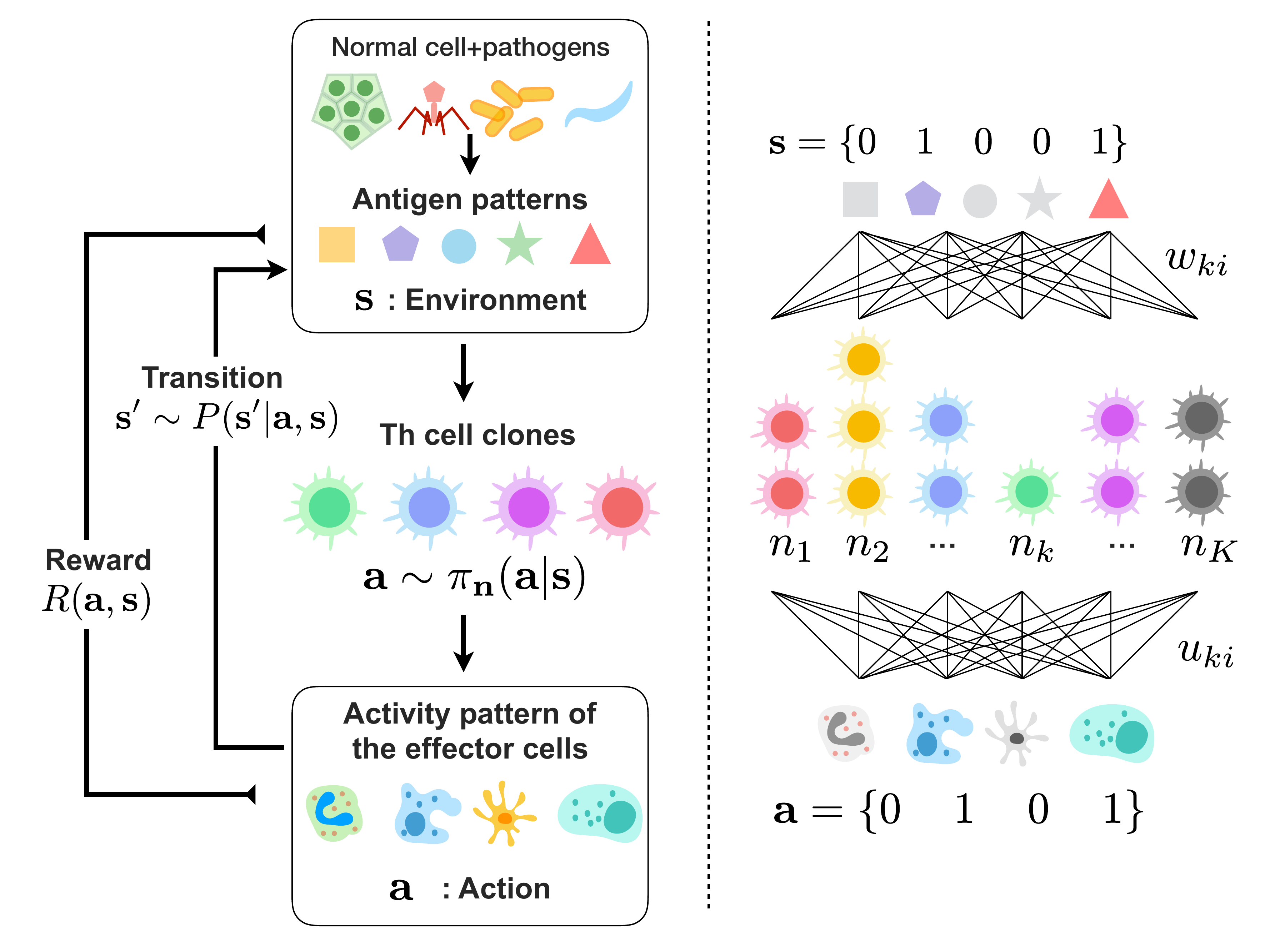}
\caption{Schematic diagram of the adaptive immune system in relation with the network-based reinforcement learning. When an infection occurs, APCs engulf the pathogens and present multiple antigens as an antigen pattern $\mathbf{s}$.
The Th cell population recognizes the antigen pattern $\mathbf{s}$ and biases the activities of the effector cells. The stochastic mapping $\pi_{\mathbf{n}}(\mathbf{a}|\mathbf{s})$ from $\mathbf{s}$ to $\mathbf{a}$ is regarded as the policy of the system parameterized by the abundance of the Th clones $\mathbf{n}$. 
The effectiveness of the pattern of the effector activities $\mathbf{a}$ to the infection is represented as the reward $R(\mathbf{s},\mathbf{a})$.}
\label{fig:mdp_schematic}
\end{figure}

At the algorithmic level, the system has to find a better way to bias the activities of  the effector cells to the infected pathogens. 
For example, activating T killer cells is effective in coping with virus-infected cells but not with bacteria. 
The Th cells coordinate this process; they obtain the information of the infected pathogens from the pattern of antigens presented by antigen presenting cells (APCs).
Then, the Th cells regulate the activities of groups of effector cells by secreting different kinds of cytokines. 
As a network, the Th cell population constitutes the middle layer between the pattern of antigens and that of the activated effector cells (Fig. \ref{fig:mdp_schematic}). 
By following the recent advancements in the applications of neural networks for solving RL problems \cite{mnih2013playing, silver2016mastering}, we derive the learning dynamics of the Th cell population, which corresponds to the algorithm to achieve the goal formulated at the computational level. 
The derived learning dynamics has the form of a replicator equation, which can be interpreted at the implementation level as the clonal selection of the Th cells in response to antigen presentation. 
The derived learning rule also contains the terms that work as feedback from the effector cells to the Th cells.
These results provide us with fruitful insights on the potential roles of molecular and cellular components for learning in real immune systems. 
The simulations of MDP with the derived learning dynamics demonstrate that the clone size distributions of the Th cell population after learning can show properties that are qualitatively consistent with those observed experimentally.

\TJKK{It should be noted that our formulation is not intended to account for all the details of immunity but to highlight the learning aspect of immunity.While we focus primarily on Th cells, we also discuss how other components and constraints of the immune system can be incorporated. Our approach can also complement more mechanistic investigations of the dynamics and regulation of immune responses by suggesting the functional roles of such dynamics at the computational and algorithmic levels.}

\section{Model}

\subsection{Framing adaptive immune response and learning as reinforcement learning}

Upon infection by a pathogen, the innate immune responses are initiated. Subsequently, the APCs that engulf the pathogen start presenting peptide-fragments of the pathogen (antigens) to the Th cells. 
In general, multiple peptide fragments are derived from a pathogen and their pattern works as a fingerprint of the pathogen. 
Let $N$ be the number of different types of antigens and $\mathbf{s} \in \{ 0, 1\}^N$ be a pattern of the antigens; $s_i = 1$ and $s_i = 0$ indicate the presence and absence of the $i$th type of antigen, respectively (Fig. \ref{fig:mdp_schematic}). 
An antigen pattern $\mathbf{s} = \{0, 1, 1, 1\}$, for example, indicates that all but the first type of antigens exist among the four. This antigen pattern conveys information about the infected pathogens to the Th cells.

Upon receiving the information, the Th cell population secretes a pattern of cytokines, which may differ depending on the activities of the Th cells induced by the antigen pattern $\mathbf{s}$. 
In turn, depending on the cytokine pattern, different groups of effector cells, which include B cells, T killer cells, macrophages, etc., are activated or deactivated, constitute the response to the pathogen. 
It should be noted that these effector cells are different from the effector T cells.
Let $M$ be the number of different types of effector cells and $\mathbf{a} \in \{0, 1\}^M$ be an activation pattern of the effector cells. 

Here, $a_j = 1$ and $a_j = 0$ indicate the activation and inactivation of the $j$th type of effector cells, respectively (Fig. \ref{fig:mdp_schematic}). 
An activation pattern $\mathbf{a}  = \{1, 1, 0\}$, for example, implies that the first and second types of effector cells are activated while the third one is inactivated. The Th cell population can bias the activation pattern $\mathbf{a}$ of the effector cells based on the information of the antigen pattern $\mathbf{s}$. 
We express this role of the Th cell population by a stochastic transition probability $\pi(\mathbf{a} | \mathbf{s})$ that determines which activation pattern $\mathbf{a}$ is likely to be realized when the Th cells are exposed to the antigen pattern $\mathbf{s}$. 
We call this conditional probability distribution $\pi$ the policy of the Th cell population.

Patterns of the activated effector cells have different influences on the antigen patterns. 
If the activated effector cells are effective for the pathogen, the types of antigens that are specific to the pathogen should disappear from the antigen pattern with a high probability. 
Otherwise, the antigen pattern $\mathbf{s}$ may not change much. 
We express this stochastic transition of $\mathbf{s}$ with a transition probability $P(\mathbf{s}' | \mathbf{s}, \mathbf{a})$, where $\mathbf{s}$ and $\mathbf{s}'$ denote the antigen patterns before and after being exposed to $\mathbf{a}$ pattern of effector cells, respectively. 
It should be noted that this transition law itself is physically determined by the nature of the effector cells and pathogens and that each transition may not necessarily be dependent on its immunological effectiveness. 
Immunological effectiveness is a vague but important factor with which Th cells can learn a better policy $\pi$ to induce a better activity pattern $\mathbf{a}$ for a given antigen pattern $\mathbf{s}$. 
\TJKK{The immunological effectiveness of action $\mathbf{a}$ for antigen pattern $\mathbf{s}$ is modeled here using a reward function $R(\mathbf{s}, \mathbf{a}) \in [0, \infty)$ (Fig. \ref{fig:mdp_schematic}). }
The reward function can be a complicated function of the antigen and activation patterns in general, but it can be presumed to take a large value if the activity pattern of the effector cells is effective for the current state; otherwise, it takes a low value. 
The details and a biological counterpart of this reward signal will be discussed in a later section. 

In summary, the learning dynamics of the immune cell population are modeled by the following five components.
\begin{itemize}
    \item A set of antigen patterns $\mathcal{S} \subset \{ 0, 1\}^N$,
    \item A set of activity patterns of the effector cells $\mathcal{A} \subset \{ 0, 1\}^M$,
    \item Transition probability $P(\mathbf{s}' | \mathbf{s}, \mathbf{a})$,
    \item Reward function $R(\mathbf{s},\mathbf{a}) \in \mathbb{R}$,
    \item Policy of Th cell population $\pi(\mathbf{a}|\mathbf{s})$.
\end{itemize}
In the terminology of MDP, the first four components correspond to a set of states, a set of actions, transition probability, and reward, respectively.

\TJKK{By optimizing the policy $\pi(\mathbf{a}|\mathbf{s})$ via interactions with pathogens, the immune system can adaptively respond to infections.
The optimal policy $\pi^{\dagger}$ is characterized as the policy that maximizes the expected cumulative reward $J[\pi] := \mathbb{E}\left[\sum_{t=0}^{\infty}\gamma^{t}R^{\pi}(\mathbf{s}_{t},\mathbf{a}_{t})\right]$ as $\pi^{\dagger} :=\arg \max_{\pi} J[\pi]$, where 
\begin{align*}
R^{\pi}(\mathbf{s}_{t},\mathbf{a}_{t})=R(\mathbf{s}_{t},\mathbf{a}_{t})-\frac{1}{\beta}\log \frac{\pi(\mathbf{a}_{t}|\mathbf{s}_{t})}{\pi_{0}(\mathbf{a}_{t}|\mathbf{s}_{t})}.
\end{align*}
The additional term $\frac{1}{\beta}\log \pi/\pi_{0}$ represents the cost of control to bias the activity of the effector cells from their intrinsic behavior $\pi_{0}$ to $\pi$. 
$\beta \in (0, \infty)$ is a scaling parameter of the cost. 
Because of the functional form of the control cost, this formulation is also recognized as an entropy-regularized reinforcement learning (erRL)\cite{Neu2017arXiv}. 
For simplicity, we assume here that $\pi_{0}$ is uniform.
The optimal policy can be explicitly represented as 
\begin{align*}
\pi^{\dagger}(\mathbf{a}|\mathbf{s}) = \frac{\exp[ \beta Q^{\dagger} (\mathbf{a},\mathbf{s})]}{\sum_{\mathbf{a}} \exp [ \beta Q^{\dagger} (\mathbf{a},\mathbf{s})]},
\end{align*}
where the optimal Q function $Q^{\dagger}$ is defined as $Q^{\dagger} :=\max_{\pi} Q(\mathbf{s},\mathbf{a})$ and 
\begin{align}
    Q(\mathbf{s},\mathbf{a}) := \mathbb{E}\left[R(\mathbf{s}_{0},\mathbf{a}_{0}) + \sum_{t=1}^{\infty}\gamma^{t}R^{\pi}(\mathbf{s}_{t},\mathbf{a}_{t})\middle|
    \begin{array}{rl}\mathbf{s}_{0}&=\mathbf{s},\\
    \mathbf{a}_{0} &=\mathbf{a}
    \end{array}\right].\label{eq:Qfunc0}
\end{align}
While the optimal policy $\pi^{\dagger}$ is obtained theoretically, it is not clear how the immune system can implement it. }
Moreover, because the immune system does not have perfect information on $P(\mathbf{s}' | \mathbf{s}, \mathbf{a})$ and $R(\mathbf{s},\mathbf{a})$ \textit{a priori}, 
the optimal policy should be learned via interactions with pathogens \textit{a posteriori}.

\subsection{Implementation of policy by T helper cell population}

Each T helper cell can be characterized by its T cell receptor (TCR) and the types of cytokines secreted, which roughly correspond to the phenotypic subtypes of the Th cells, \textit{e.g.}, Th1, Th2\cite{murphy2016janeway,Ellmeier2018NatRevImmunol}. 
Let $K$ be the number of different Th clones classified according to these criteria, and $n_k$ be the population size of the $k$th clones. 
Each clone interacts with the $i$th antigen with a different strength, which is determined by the affinity of the TCR of the clones to the antigen and also by how the antigen is presented. 
Because each clone has a unique TCR, the interaction strength $w_{ki} \in \mathbb{R}$ of the $k$th clone to the $i$th antigen is the same amongst all cells of the $k$-th clone. 
When the antigen pattern is $\mathbf{s}$, each Th cell of type $k$ is supposed to receive stimulation $w_{ki} s_i$ from the $i$th antigen. 
The total stimulation that each Th cell of type $k$ receives becomes $\Sigma_{i} w_{ki} s_i$. 
A Th cell of type $k$ activates itself, and its activity, $h_k(\mathbf{s})$, is assumed to be dependent sigmoidally on the strength of the stimulation as
\begin{equation}
    h_k(\mathbf{s}) = \sigma\left(\sum_{i=1}^N w_{ki} s_i \right),
\end{equation}
where $\sigma$ is the sigmoid function and $\sigma(x) = 1/ (1+\exp(-x))$. 
The activity $h_k$ becomes either $1$ or $0$ when the cell is fully activated or deactivated, respectively. 
Such monotonous and sigmoidal dependency is consistent with several experimental observations \cite{murphy2016janeway}.

Depending on their activities, the Th cells release cytokines, which in turn bias the activities of the effector cells. 
All Th cells of type $k$ are assumed to release the same types  of cytokines, because they belong to the same effector subtype. 
A stimulus to the $j$th type of effector cells via cytokines released from a Th cell of the $k$th type is expressed as $\beta u_{jk} h_k$, where $h_k$ is the activity of the $k$th clone, $u_{jk} \in \mathbb{R}$ defines the strength and sign of the stimulus, and $\beta \in (0, \infty)$ is a global scaling parameter.  

The integral stimuli received by the $j$th effector cells from all Th cells is then represented as $\Sigma_k n_k h_k u_{jk}$. 
In response to this integral stimuli, the probability that the $j$th type of effector cells is activated ( $a_j = 1$) or deactivated ($a_j = 0$) is modulated according to the following conditional probability:
\begin{equation}
    p(a_j=1 | \mathbf{s}) = \sigma \left(\sum_{k=1}^K \beta u_{jk} n_k h_k(\mathbf{s})\right),
\end{equation}
where we suppose that the cytokines are the major bias factors of the activities of the effector cells by Th cells.
\TJKK{The positive and negative biases may be associated with inflammatory and anti-inflammatory cytokines, respectively.}
It should be noted that, in this formulation, the effector cells have the autonomous ability to be activated or deactivated, which is biased by the signal from the Th cells. While we can consider that such autonomous activity is directly modulated by the pathogens, we assume that it is independent for simplicity and for focusing mainly on the roles of the Th cells.

Therefore, the Th cell population translates the antigen pattern $\mathbf{s}$ that it receives into an activation pattern of the effector cells $\mathbf{a}$ with a probability $\pi_{\mathbf{n}}$:
\begin{eqnarray}
    \pi_{\mathbf{n}}(\mathbf{a}|\mathbf{s}) &=& \prod_{j=1}^M p(a_j|\mathbf{s}) = \frac{\exp(\beta \tilde{Q}_{\mathbf{n}}(\mathbf{s},\mathbf{a}))}{\sum_{\mathbf{a} \in \mathcal{A}} \exp(\beta \tilde{Q}_{\mathbf{n}}(\mathbf{s},\mathbf{a}))} , \label{eq:policy}
\end{eqnarray}
where $\tilde{Q}_{\mathbf{n}}$ is defined as
\begin{equation}
    \tilde{Q}_{\mathbf{n}}(\mathbf{s},\mathbf{a}) = \sum_{k=1}^K n_k h_k(s) \sum_{j=1}^M u_{jk} a_j .\label{eq:Qfunc}
\end{equation}
\TJKK{This conditional probability is the policy of the immune system implemented by the Th cell population, and the role of the Th cell population is to update the policy over time  by modulating the clone size distribution $\mathbf{n}$ so as to make $\pi_{\mathbf{n}}$ closer to $\pi^{\dagger}$ in order to receive a greater reward.}

\subsection{Learning dynamics of the Th cell population}

Similarly to $\pi^{\dagger}$, the policy $\pi_{\mathbf{n}}(\mathbf{a}|\mathbf{s})$ is represented in the form of Boltzmann distribution with respect to $\mathbf{a}$ in which $\tilde{Q}_{\mathbf{n}}(\mathbf{s}, \mathbf{a})$ and $\beta$ are the negative energy and the global scaling parameter, respectively. 
Because of this form, the policy $\pi$ is likely to select an activity pattern of the effector cells $\mathbf{a}$ with greater value of $\tilde{Q}_{\mathbf{n}}(\mathbf{s}, \mathbf{a})$ than the others.
If $\tilde{Q}_{\mathbf{n}}(\mathbf{s}, \mathbf{a})$ represents the value of choosing $\mathbf{a}$ in response to $\mathbf{s}$, the policy $\pi_{\mathbf{n}}(\mathbf{a}|\mathbf{s})$ implemented by the Th cell population can be interpreted as a strategy to choose the activity pattern of a higher $\tilde{Q}_{\mathbf{n}}(\mathbf{s}, \mathbf{a})$ value with higher probability than the others.
In terms of maximizing the reward, the immune system should select the activity pattern of the effector cells $\mathbf{a}$ that returns a higher reward $R(\mathbf{s},\mathbf{a})$ in response to an antigen pattern $\mathbf{s}$.
Therefore, intuitively, the policy $\pi_{\mathbf{n}}(\mathbf{a}|\mathbf{s})$ becomes better when $\tilde{Q}_{\mathbf{n}}(\mathbf{s},\mathbf{a})$ has been updated to represent the reward $R(\mathbf{s},\mathbf{a})$ more faithfully.
\TJKK{This intuitive interpretation can be rationalized by considering $\gamma=0$ for \eqref{eq:Qfunc0}. }
The policy in the Boltzmann form of \eqref{eq:policy} is shown to be optimal when $\tilde{Q}_{\mathbf{n}}(\mathbf{s},\mathbf{a})$ becomes identical to the optimal $Q^{\dagger}$ function, which is equal to $R(\mathbf{s},\mathbf{a})$ in this case.
Therefore, optimizing $\pi_{\mathbf{n}}(\mathbf{a}|\mathbf{s})$ is equivalent to learning $\tilde{Q}_{\mathbf{n}}(\mathbf{s},\mathbf{a})$, which is an estimate of the reward function $R(\mathbf{s},\mathbf{a})$, from the past experiences of interactions with pathogens \cite{Neu2017arXiv}.

Therefore, the learning dynamics can be reduced to updating $\tilde{Q}_n(\mathbf{s},\mathbf{a})$ to be closer to the reward function $R(\mathbf{s},\mathbf{a})$ than before by modulating the clone size distribution $\mathbf{n}$. 
One way to derive such an update rule is to minimize the following cost function with respect to the parameter $\mathbf{n}$ in $\tilde{Q}_{\mathbf{n}}(\mathbf{s},\mathbf{a})$ for each episode $\mathbf{s}$,$\mathbf{a}$, and $r=R(\mathbf{s},\mathbf{a})$:
\begin{equation}
    L_{\mathbf{n}}(\mathbf{s},\mathbf{a}) = \frac{1}{2}\left(r - \tilde{Q}_{\mathbf{n}}(\mathbf{s},\mathbf{a})\right)^2.
\end{equation}
$\tilde{Q}_{\mathbf{n}}(\mathbf{s},\mathbf{a})=R(\mathbf{s},\mathbf{a})$ is achieved when this cost function $L_{\mathbf{n}}$ takes the minimum value $0$ for all pairs of $(\mathbf{s},\mathbf{a})$. 
If a biological learning system were equipped with a versatile memory that could store the experienced rewards for all pairs of $(\mathbf{s},\mathbf{a})$, $\tilde{Q}_{\mathbf{n}}(\mathbf{s},\mathbf{a})=R(\mathbf{s},\mathbf{a})$ would seem to be achieved trivially.
However, such implementation is not feasible both biologically and computationally.
Storing such information requires a large memory, the capacity of which is of the order of $2^{M+N}$. 
Moreover, owing to the lack of generalization in this implementation, i.e., the experienced rewards are not exploited to infer the reward of the not-yet-experienced pairs of $(\mathbf{s},\mathbf{a})$, the system needs an extraordinarily long time to experience all pairs.
Recent advancements in \TJKK{network-based} reinforcement learning have demonstrated that the implementation of the Q function by a neural network is efficient in terms of both memory usage and generalization \cite{mnih2013playing}.

The Th cells form a network similar to  neural network (Fig. \ref{fig:mdp_schematic}) \TJKK{and can potentially approximate $R(\mathbf{s},\mathbf{a})$ in the form of $\tilde{Q}_{\mathbf{n}}$ in \eqref{eq:Qfunc}.}
However, experimental evidence suggests that Th cells realize learning mainly by adjusting their clone size distribution $\mathbf{n}$ and that the other parameters such as the weights $\mathbf{w}$ of the Th clone-antigen interactions may not be changed. 
This is in sharp contrast to the case of neural networks in which the interaction weights can be directly modulated to achieve learning. Thus, it is not obvious whether learning can be achieved in the immune system only by adjusting $\mathbf{n}$.

In addition, a simple update of $\mathbf{n}$ along the gradient of the cost function, $\nabla_{\mathbf{n}} L_{\mathbf{n}}(\mathbf{s}, \mathbf{a})$, may not necessarily be relevant biologically because the learning dynamics should satisfy the \TJKK{invariance constraint with respect to the subdivision of a clone} that is imposed by the interpretation of $\mathbf{n}$ as the clone size distribution.
Suppose that the $k$th clone accommodates $n_{k}$ cells and is subdivided into two sub-clones, $k_{1}$ and $k_{2}$, as $n_k = n_{k_{1}} + n_{k_{2}}$. 
Such a subdivision should not change the learning dynamics as long as the two subclones have the same properties as those before the subdivision.
To satisfy \TJKK{this invariance rule}, the following metric of the parameter space $\mathbf{n}$ should be considered:
\begin{equation}
    g_{ij}({\bf n}) = \delta_{ij} / n_i.
\end{equation}
With this metric, the appropriate gradient can be derived as
\begin{equation}
 \{\nabla_{\mathbf{n}}^{g} L_n(s, a)\}_{k}=   \sum_m g^{-1}_{km} \frac{\partial L_n(\mathbf{s}, \mathbf{a})}{\partial n_m } = n_k \frac{\partial L_n(\mathbf{s}, \mathbf{a})} {\partial n_k} ,
\end{equation}
which is the natural gradient in the parameter space $\mathbf{n}$ \cite{amari1998natural}. 
Thus, when the Th population with a clone size distribution $\mathbf{n}(t)$ at time $t$ experiences an antigen pattern $\mathbf{s}(t)$ and an activation pattern $\mathbf{a}(t)$ of the effector cells, the update rule of $\mathbf{n}$ can be derived in the form of a replicator equation as
\begin{align}
    n_k(t+1) &= n_k(t) + \alpha \,n_k(t) \lambda_k(t) , \label{eq:learn_eq}
\end{align}
where the positive constant $\alpha$ is the learning rate, 
\begin{align}
\lambda_k(t) &:= \frac{\partial L_n(\mathbf{s}, \mathbf{a}, r) }{ \partial n_k}\\
&= \left[r(t) - \tilde{Q}(\mathbf{s}(t), \mathbf{a}(t))\right] h_k(\mathbf{s}(t)) \sum \nolimits_j u_{jk} a_j(t), \label{eq:fitness}
\end{align}
and $r(t):=R(\mathbf{s}(t), \mathbf{a}(t))$.
This rule of learning dynamics is similar to that of SARSA or Q learning with a linear functional approximation \cite{sutton1998reinforcement}.
The details of the dynamics show that the Th cell population can learn an effective response to each pathogen if each clone of type $k$ proliferates or dies by following the growth rate $\lambda_k$. 
The self-replicative nature of the dynamics originates from the metric $g_{ij}({\bf n})$, whereas the functional form of the growth rate $\lambda_k(t)$ is determined by the gradient of $L_n(\mathbf{s}, \mathbf{a}, r)$.
Therefore, the self-replicative nature of \eqref{eq:learn_eq} is invariant to changes in the details as long as the Th clone size works as the learning parameter.

\subsection{Biological interpretation of learning dynamics}

The derived learning dynamics can be interpreted biologically by introducing the following decomposition of the growth rate $\lambda_k$ of the $k$th clones:
\begin{align}
    \lambda_k(\mathbf{s}, \mathbf{a}, r) &= f_k(\mathbf{s}, \mathbf{a}) \left[r  - \tilde{Q}_{\mathbf{n}}(\mathbf{s}, \mathbf{a})\right],
\end{align}
where
\begin{align}
    \tilde{Q}_{\mathbf{n}}(\mathbf{s}, \mathbf{a}) &= \sum \nolimits_l n_l f_l(\mathbf{s}, \mathbf{a}),& f_k(\mathbf{s}, \mathbf{a}) &:= h_k(\mathbf{s}) \sum \nolimits_j u_{jk} a_j .
\end{align}
$\left[r  - \tilde{Q}_{\mathbf{n}}(\mathbf{s}, \mathbf{a})\right]$ is common to all clones and can be interpreted as a global signal to all Th cells. 
In contrast, $f_k(\mathbf{s}, \mathbf{a})$ determines the clone-specific sensitivity to that signal.
\TJKK{Further, $h_k(\mathbf{s})$ in $f_k(\mathbf{s}, \mathbf{a})$ is the antigen-dependent activity of the $k$th clones, whereas $\sum \nolimits_j u_{jk} a_j$ is the feedback from the active effector groups.
This indicates that the $k$th clones have a high sensitivity to the global signal when they receive a strong antigenic signal from the current antigen pattern and also have  feedbacks from the effector groups\cite{Kalinski2005NatRevImmunol,annunziato20153}.}
\TJKK{Such feedback requires local interactions between the Th cells and the effector groups as cytokines mediate the paracrine communications.
Biologically, the Th cell population is known to proliferate only when exposed to both signals, namely the stimulus to TCR and the co-stimulus from APCs or innate immune cells \cite{chen2013molecular, murphy2016janeway}.
There are additional pieces of evidence that indicate interactions between the signals from TCR and cytokines that might be released by the effector groups \cite{voisinne2015t, tkach2014t}.}
\TJKK{Moreover, pro-inflammatory cytokines are secreted by Myeloid cells to activate naive Th cells\cite{kopf2010averting}.}
It should be noted that $f_k(\mathbf{s}, \mathbf{a})$ can be negative if the $k$th clone has an inhibitory effect on certain effector groups. 
Such a situation might be related to the activation-induced cell death (AICD) of T cells \cite{maher2002activation, huang1999activation}. 
T helper cells express Fas ligands as they are activated. 
While these Fas ligands have no significant effect on inactive T cells, they can induce apoptosis on the activated T cells, which is considered as a mechanism of immune tolerance\cite{maher2002activation}.
Additionally, the negative $f_k(\mathbf{s}, \mathbf{a})$ may be interpreted as the action of anti-inflammatory cytokines.
Our result suggests that these positive and negative back-propagating controls may be responsible for modulating the relative contributions of the T cell clones depending on the consistency between the activities of the Th cells and the effector groups.
Moreover, the derived learning rule indicates that such local modulations are not sufficient for learning because the individual Th clones are blind to whether their activities and those of the induced effector groups have actual immunological impact. 
The global signal $\left[r  - \tilde{Q}_{\mathbf{n}}(\mathbf{s}, \mathbf{a})\right]$ is indispensable for conveying that information and might be associated with the damage signal\cite{pradeu2012frontimmunol} or physiological conditions of the body such as temperature or endocrine signals.
While the biological interpretations of the local and global signals are not decisive,  the learning rule highlights the roles and necessity of these two different types of interactions among immune cells for achieving appropriate learning. 

\begin{figure}[tbhp]
\centering
\includegraphics[width=0.98\linewidth]{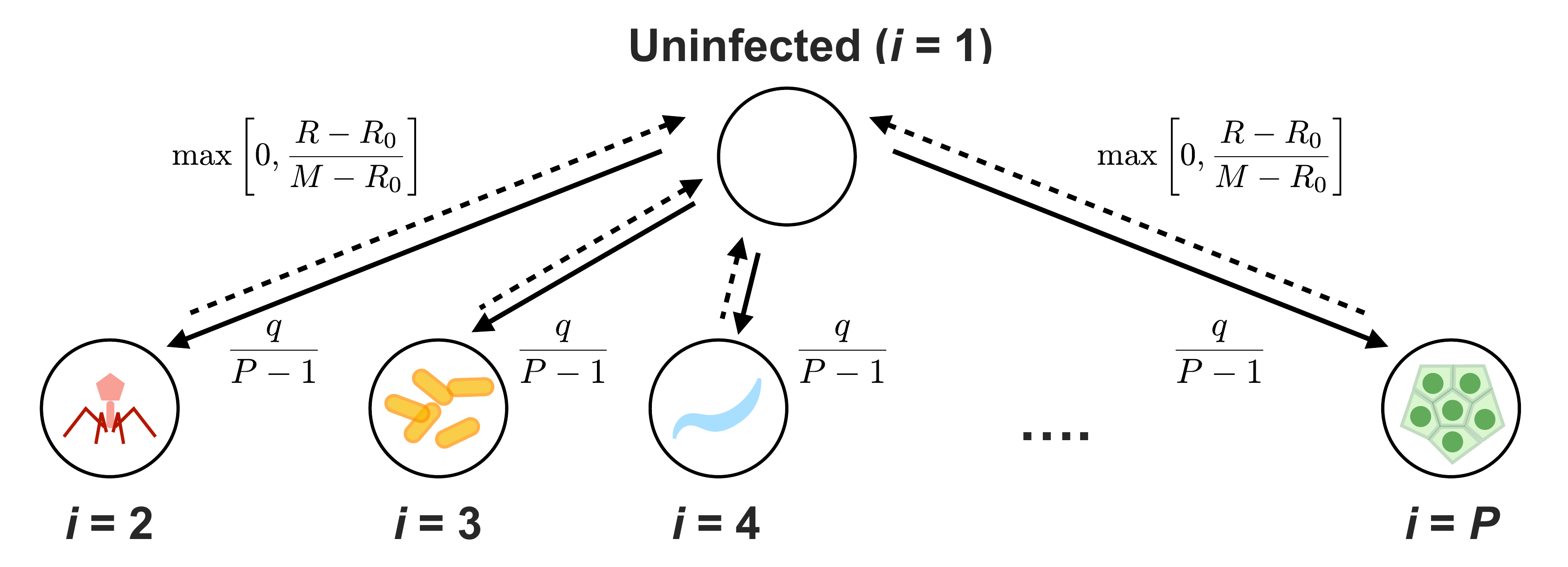}
\caption{Schematic diagram of the stochastic transition dynamics of the infected and uninfected states (the environment in MDP) defined by Eq. \ref{eq:transition_prob}. In all simulations, $q=0.9$ and $R_{0}=M/2$ are used.}
\label{fig:PathogenDynamics}
\end{figure}

\section{Numerical simulations and clone size distribution after learning}

We conducted simulations to confirm that the derived dynamics could actually learn because the steepest descent method does not always guarantee their resulting performance. 
We also investigated stationary clone size distributions of the Th cell population to draw insights on the behavior of an appropriately trained learning system after learning and to check the consistency with experimentally observed results.
Because individual antigens, Th clones, and effector types are modeled explicitly in our formulation, simulations with realistic parameter values are prohibitive. \TJKK{For example, the varieties of Th clones, $K$, can be of the order of $10^{6}$ or more \cite{Laydon2015}. While we lack a reliable estimate, the variety of antigens $N$ can be of a similar order as that of $K$ because each antigen is characterized by the peptide sequence of length $8-9$. The number of effector types $M$ should be much of a smaller order because the effector types are determined by genetically encoded cell types and their phenotypic states.}
To circumvent this difficulty, we instead focus on the general properties of the learning dynamics and stationary distribution for a much smaller and tractable parameter set and investigate  the scaling property with regard to the change in the parameter values.

We assume that there are $N$ distinct antigens and $P$ different  antigen patterns $\{ \mathbf{s}^1, \mathbf{s}^2, \cdots, \mathbf{s}^{P}\}$ representing uninfected and infected states.
$\mathbf{s}^1$ corresponds to the antigen patterns of the uninfected state and $\mathbf{s}^{i}$ corresponds to that of the $i$th pathogen. 
We also assume that there are $M$ different types of effector cells and the associated $P$ possible activity patterns $\{ \mathbf{a}^1, \mathbf{a}^2, \cdots, \mathbf{a}^{P} \}$, each of which represents the most effective activity pattern of the effector cells to the corresponding pathogen.
While the possible antigen and activity patterns, in principle, depend on the kind of pathogens considered, such detailed information is experimentally available only for very exceptional cases \cite{Graham2018}.
Thus, both antigen and activity patterns are generated randomly by following the maximum entropy principle under non-informative situations. 
For each pair of $i, j \in \{1, \cdots, P\}$, the reward function $R(\mathbf{s},\mathbf{a})$ is determined as
\begin{equation}
    R(\mathbf{s}^{i},\mathbf{a}) = M - \mathrm{ham}(\mathbf{a}^i, \mathbf{a}), \label{eq:reward}
\end{equation}
where $\mathbf{a}^i$ is the most effective activity pattern for the antigen pattern $\mathbf{s}^{i}$ and $\mathrm{ham}(\mathbf{a}, \mathbf{a}')$ is the hamming distance between two binary vectors $\mathbf{a}$ and $\mathbf{a}'$. 
This functional form indicates that the immune system receives the highest reward $M$ when the activity pattern $\mathbf{a}$ matches the most effective one for the antigen pattern $\mathbf{s}^{i}$.
If $\mathbf{a}$ deviates from the most effective $\mathbf{a}^{i}$, the immune system experiences a loss of reward by the deviation $\mathrm{ham}(\mathbf{a}^i, \mathbf{a})$.
Because $M$ is the maximum hamming distance for a pair of vectors with length $M$, $R(\mathbf{s},\mathbf{a})$ is always positive for any pair of antigen and activity patterns.

The transition probability $P(\mathbf{s}' | \mathbf{s}, \mathbf{a})$ is selected as
\begin{align}
    P(\mathbf{s}^i | \mathbf{s}^j, \mathbf{a}) =
    \begin{cases}
    1-q & \mbox{if $j=1$ \& $i = 1$}\\
    q/(P-1) & \mbox{if $j=1$ \& $i \neq 1$}\\
    \min\left[1,\frac{M- R(\mathbf{s}^{i},\mathbf{a})}{M-R_{0}}\right] & \mbox{if $j\neq 1$ \& $i = j$}\\
    \max\left[0, \frac{R(\mathbf{s}^{i},\mathbf{a})-R_{0}}{M-R_{0}}\right] & \mbox{if $j\neq 1$ \& $i = 1$}\\
    0 & \mbox{otherwise}
    \end{cases}.
 \label{eq:transition_prob}
\end{align}
The transition probability represents the dynamics depicted in Fig.\ref{fig:PathogenDynamics}. 
Here, $q$ is the probability to be infected by a pathogen, which is randomly chosen from $P-1$ pathogens equally. 
If infected, the pathogen is swept out with the probability $\max\left[0, \frac{R(\mathbf{s}^{i},\mathbf{a})-R_{0}}{M-R_{0}}\right]$. 
This means that if the reward is its maximum $M$, the pathogen is eliminated with probability $1$. 
On the other hand, the pathogen cannot be removed if the reward is less than or equal to the threshold $R_{0}$. 
As the reward increases from $R_{0}$, the chance of recovery increases linearly. 
If $R_{0}$ is small, the pathogen can be removed with a certain probability even without effective control from the Th cells due to the intrinsic ability of the effector cells.
If $R_{0}$ is very close to the maximum reward, the learning is typically hampered by being trapped in one of the infected state.
This suggests that the innate ability to recover from infections is a requisite for adaptive learning.
\TJKK{We can consider more complicated situations by extending the space of $\mathbf{s}$. For example, by assigning several states of $\mathbf{s}$ to each pathogen, we can represent the stages of  infection, the transitions between which are dependent on the action of the immune system. 
Such a model can be effectively used to analyze more complicated infections such as chronic ones.
We may also model adversarial ones where the next infection is dependent on the current infection due to the co-evolution of pathogens.}

\begin{figure}[htbp]
\centering
\includegraphics[width=0.98\linewidth]{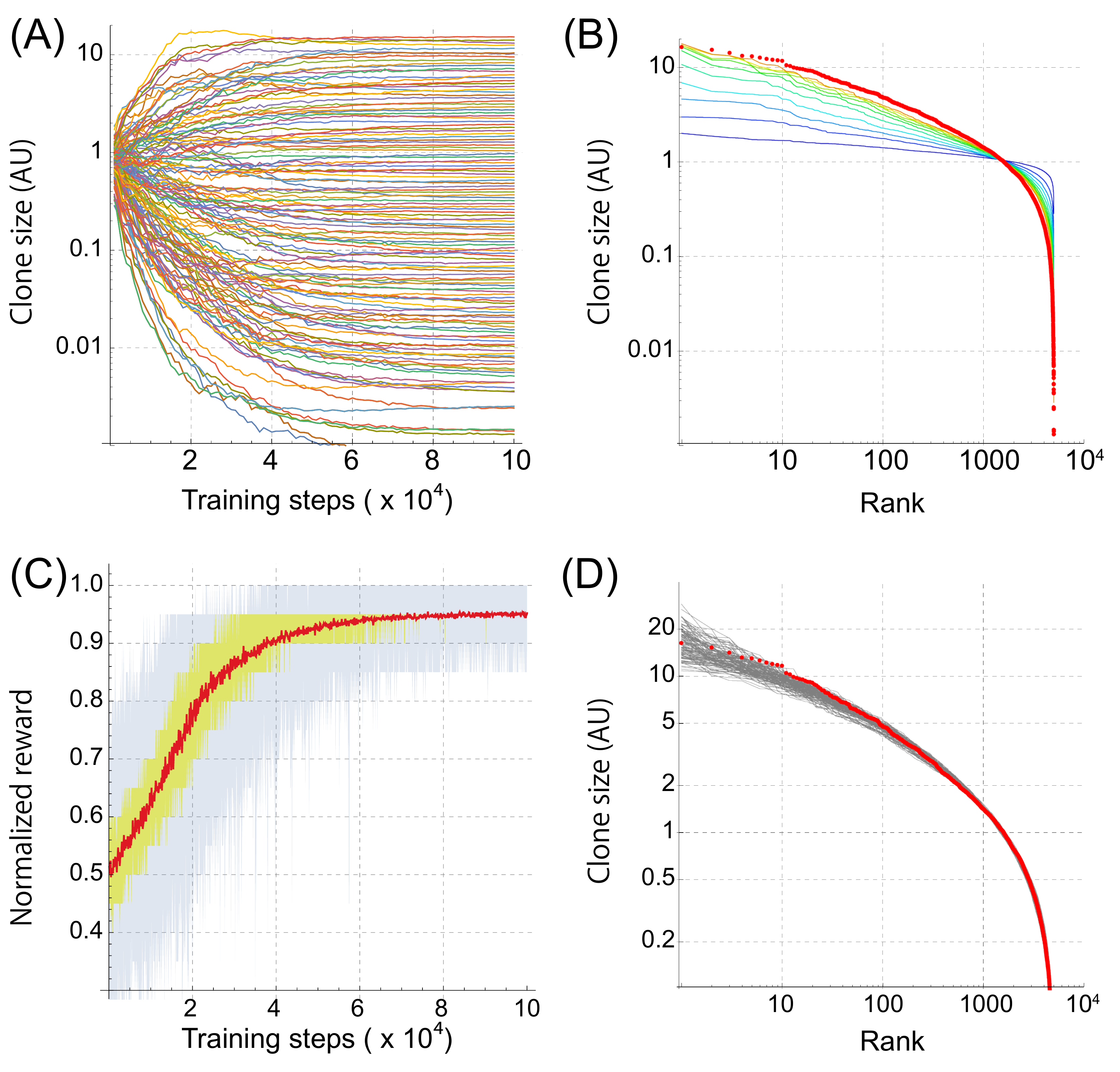}
\caption{
(A) Trajectories of the Th clone size $\mathbf{n}$ along a learning trial. 
The parameter values are $N=100$, $K=5,000$, $M=20$, and $P=30$.
The clone size $\mathbf{n}$ is normalized by the initial abundance. 
This figure shows only $200$ trajectories sampled evenly out of $5000$ ones at the stationary state to avoid complication of the plot.
(B) The dynamics of rank-abundance distributions along the learning trial, which were calculated from the trajectories of $\mathbf{n}$ in (A).
The red circles represent the stationary distribution after the learning, and the colored curves are the transient distributions calculated at training steps from $1 \times 10^{3}$ (blue) to $28  \times 10^{3}$(yellow).
(C) Statistics of learning curves of the Th cell population. The rewards normalized by its maximum value are obtained as functions of the training step for $100$ independent learning trials. 
The red curve is the average reward, and the yellow and blue regions show the range between $25$ and $75$ percentiles of the rewards and that between minimum and maximum of the rewards at each training step, respectively.
(D) The stationary rank-abundance distributions for the $100$ independent learning trials in (C) are shown by gray curves. The red circles are the same as those in (B).
}
\label{fig:time_series_n}
\end{figure}

Finally, we suppose that the distribution of the interaction strength $w_{ki}$ of the $k$th clone to the $i$th antigen follows the normal distribution with mean $0$ and variance $\sigma^{2}_{w}$ to represent the cross-reactivity of TCRs\cite{Riley2018NatChemBiol}. 
Similarly, the effect of the stimuli on the $j$th effector cells from the $k$th clone,  $u_{jk}$, is sampled from a normal distribution with mean $0$ and variance $\sigma^{2}_{u}$, because we lack quantitative information on this parameter\cite{Kveler2018NatBiotech}.
It should be noted that the affinity of a TCR towards antigens is expected to be sparse such that a TCR reacts to a small fraction of all possible antigens. 
Even if this sparsity is considered, the following results are not affected qualitatively. 
In contrast to the training of neural networks, these interaction parameters are fixed in our model and the Th clone size, $\mathbf{n}$, is the only tunable parameter for learning.

Figures \ref{fig:time_series_n} (A) and (B) show the transient dynamics of the Th clone sizes and the rank-abundance distribution during a learning process \TJKK{starting from the uniform clone size distribution.}
We observe that the clone size distribution fluctuates transiently during the learning with switching of ranks of the clones, and $\mathbf{n}(t)$ eventually converges into a stationary distribution.
\TJKK{The early fluctuation is due to the small $\beta$ in the learning process (see Materials and Methods). 
This early fluctuation promotes exploration of the system, which might be related to the downregulation of the Th function in the early infancy periods \cite{schmiedeberg2016t}}
\TJKK{It should be noted that, in a real biological situation, the learning also starts with the Th clone distribution pre-trained in the thymus. 
Such pre-training may be optimized to facilitate and expedite subsequent learning, possibly by evading the very early exploring stage of the learning \cite{elhanati2014quantifying}.}
Figure \ref{fig:time_series_n} (C) shows the statistics of $100$ independent learning curves and their fluctuations.
The monotonous increase in the average reward demonstrates that the derived learning dynamics can actually work to obtain a greater reward over time by updating the clone size distribution $\mathbf{n}(t)$ based on previous experiences.
Figure \ref{fig:time_series_n} (D) shows the stationary rank abundance distributions for the $100$ independent learning trials.
Owing to the stochastic nature of the learning process, the rank-abundance distribution does not perfectly converge into an identical distribution; instead it fluctuates, which is prominent in the abundances of the highly ranked clones (i.e., $\mathrm{Rank}<100$) in Fig. \ref{fig:time_series_n} (D).

The simulations are conducted using a set of parameter values chosen as a representative situation in which learning is effectively achieved with minimum diversities of the antigens and Th cells. 
As shown in Fig. S3, further increase in either $N$ or $K$ does not improve the performance considerably, which indicates that the diversities of the antigens and clone types are sufficiently large under this condition.
\TJKK{In contrast, the performance starts declining if either $P$ or $M$ increases (Fig. S 3).  The decline induced by the increase in the number of pathogen types, $P$, is natural because learning more pathogens should be more difficult if the numbers of antigens and Th clone types are fixed. 
In contrast, the decline due to the increased $M$ highlights the importance of constraining possible actions for an efficient learning \cite{zahavy2018learn}.}

\begin{figure}[htbp]
\centering
\includegraphics[width=0.98\linewidth]{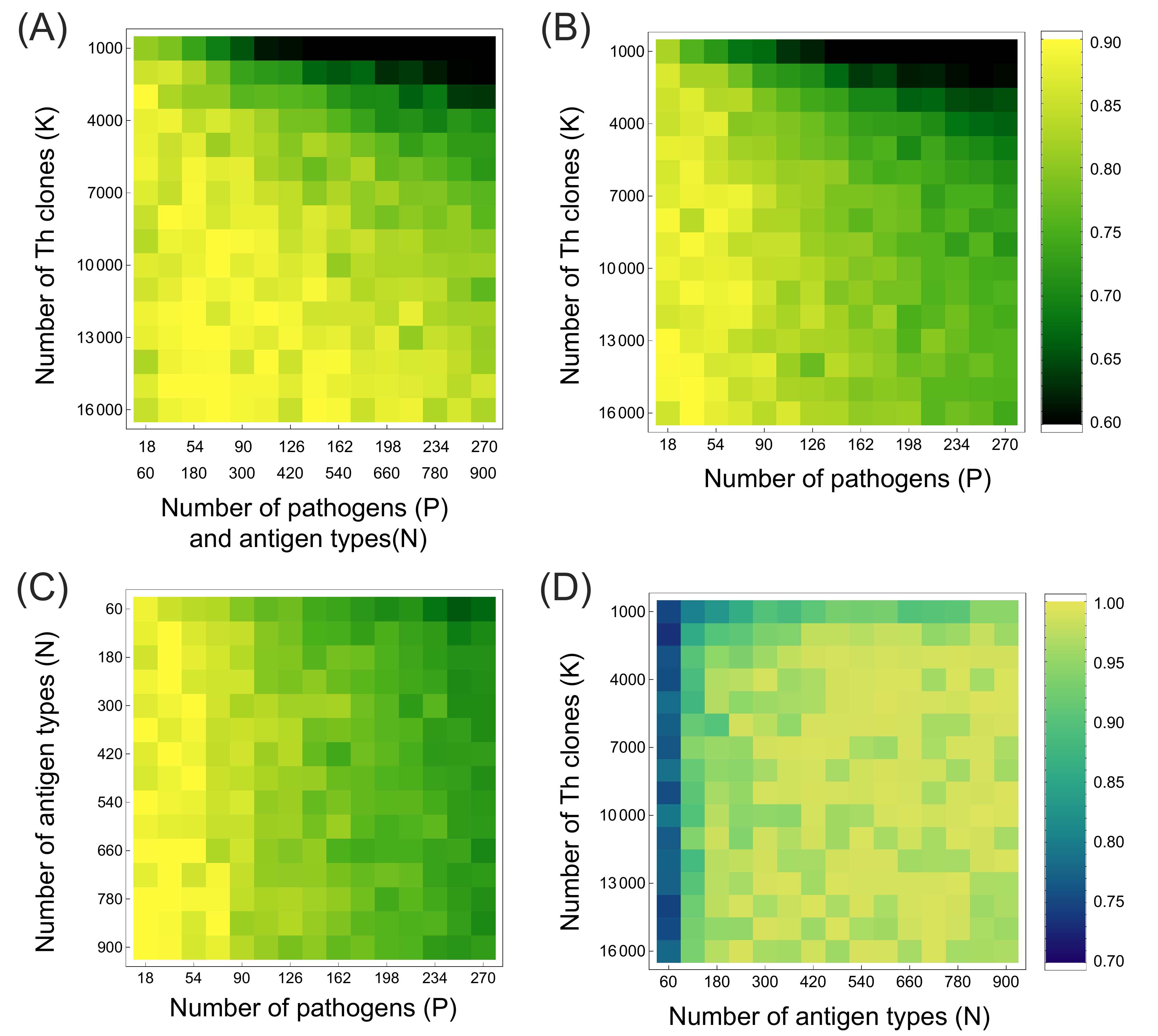}
\caption{Heatmap plots of the stationary normalized reward after learning as functions of (A) $K$ and $\{N,P\}$, (B) $K$ and $P$, (C) $N$ and $P$, and (D) $K$ and $N$. 
The other parameters are the same as those in Fig. \ref{fig:time_series_n}. 
The stationary reward was calculated as the moving average of the last $10^{4}$ steps.
(A), (B), and (C) use the same color code. 
}
\label{fig:scaling}
\end{figure}

To investigate how $N$, $K$, and $P$ can be scaled to a greater size while maintaining learning performance, we calculated the stationary reward after learning by changing $N$, $K$, and $P$ in Fig. \ref{fig:scaling}. 
Figure \ref{fig:scaling} (A) shows that the performance is approximately kept constant when $N$, $K$, and $P$ are scaled as $\xi N$, $\xi K$, and $\xi P$, where $\xi$ is the scaling parameter.
In contrast, if only either antigen diversity $N$ or Th clone diversity $K$ is increased while the other is kept constant, as in Fig. \ref{fig:scaling} (B) and (C), the increased variety of pathogens (larger $P$) cannot be handled, which indicates the importance of both diversities for learning. 
This property should be linked to the learning capacity of the network, which has been intensively analyzed for deep networks \cite{DNNC2017}.

\begin{figure}[htbp]
\centering
\includegraphics[width=0.98\linewidth]{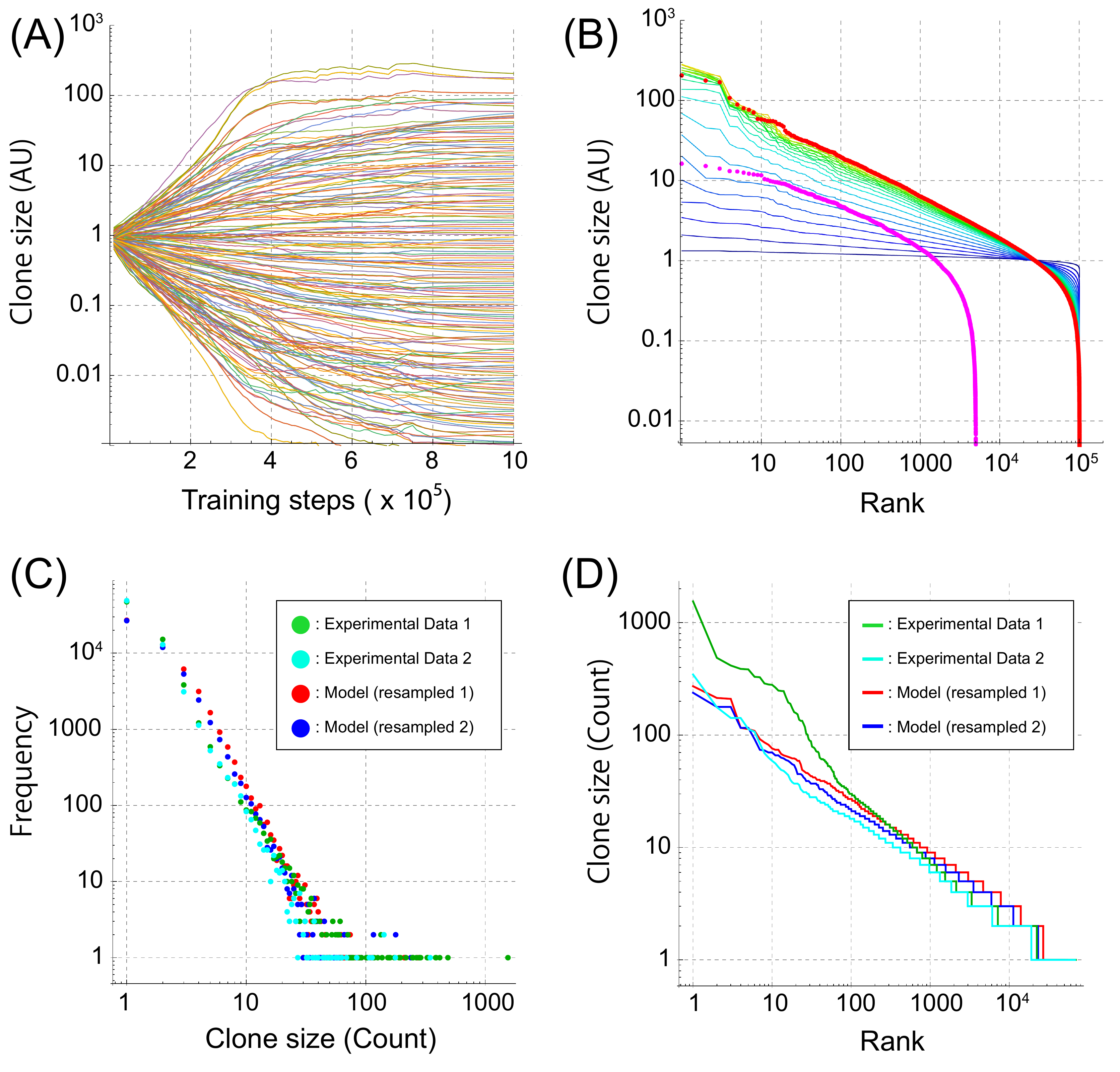}
\caption{(A) Trajectories of the Th clone size $\mathbf{n}$ along a learning trial for $K=100,000$. 
The weights, $\mathbf{w}$, are also scaled to $\mathbf{w}/10$ for comparison with the experimental data.
The other parameter values are the same as those in Fig. \ref{fig:time_series_n}.
The clone size $\mathbf{n}$ is normalized by the initial abundance. 
The figure shows only $200$ trajectories sampled evenly out of $100,000$.
(B) The dynamics of the rank-abundance distributions along the learning trial, which were calculated from the trajectories of $\mathbf{n}$ in (A).
The red circles represent the stationary distribution after the learning, and the colored curves are the transient distributions calculated at training steps from $1 \times 10^{4}$ (blue) to $77 \times 10^{4}$(yellow).
The magenta circles shown the same stationary distribution as in Fig. \ref{fig:time_series_n} (B), which is shown here for comparison.
(C, D) Clone size distributions (C) and rank-abundance distributions (D) obtained from the model and an experiment.
The green and cyan points in (C) are the clone-size distributions obtained by counting the TCR sequences of CD4+ T cells collected from the peripheral blood of two healthy human donors (HV01 and HVD4) in \cite{heather2016dynamic}, and the curves in (D) are the corresponding rank-abundance distributions.
Red and blue points in (C) are the clone-size distributions obtained by resampling the clones from the rank-distributions in (B) for the same numbers of total counts as in experiment 1 (red) and 2 (blue), and the red and blue curves in (D) are the corresponding rank-abundance distributions.}
\label{fig:shape_of_dists}
\end{figure}

Next, we investigate the effect of parameters on the shapes of the abundance distributions of the Th clones when the learning is conducted appropriately. 
For the set of parameter values in Fig. \ref{fig:time_series_n}, the rank abundance distribution in the log-log plot is relatively flat for abundant clones (from rank $10$ to $10^{3}$, approximately) but shows a sharp decline in clone size for less abundant clones (rank $>10^3$), which results in a concave distribution. 
The sharp decline in the abundance is mainly due to the limited number of Th clone types, $K$, which works as a boundary condition for the rank-abundance distribution. 
By conducting learning for a much larger $K$, (i.e., $K=10^{5}$) as in Fig. \ref{fig:shape_of_dists} (A), we observe that the relatively flat region stretches, which enhances the power-law like property of  the distribution as shown in Fig. \ref{fig:shape_of_dists} (B). 
If the total number of Th cells or observed samples is limited, very low abundant clones at the boundary are rarely observed, which can lead to a flat clone-size and rank-abundance distributions, as shown in Figs. \ref{fig:shape_of_dists} (C) and (D). 
Such flat distributions have been observed in several sequencing experiments \cite{bolkhovskaya2014assessing,dewolf2018quantifying,heather2017high,oakes2017quantitative} even though Th clone types are discriminated only by the TCR sequences in the experiments.
\TJKK{To compare the abundance distributions obtained using our model with the experimental ones, we use the TCR sequences of CD4+ T cells collected from the peripheral blood of two healthy human donors (HV01 and HVD4) reported in \cite{heather2016dynamic}.
In this dataset, molecular barcodes were added to each cDNA molecule to correct for PCR amplification and errors, which can lead to more accurate and quantitative estimates.  
The coincidence seems to be fairly good especially with Data 2 (Fig. \ref{fig:shape_of_dists} (C) and (D)).
We also note that qualitatively the same result was obtained for sparse $\{w_{ki}\}$ (Fig. 2S).}

This coincidence implies that a simple mechanism underlies the generation of power-law like distributions irrespective of the details of the dynamics.
In previous theoretical analyses, the symmetric variation in the fitness of clones was proposed as the mechanism of the power-law distribution \cite{dessalles2019heterogeneous,Desponds2016PNAS}.
Similar symmetric fitness variations are also observed in Fig. \ref{fig:shape_of_dists} (A), indicating that our model shares the same property as the previous ones under this learning condition.
The next question is the mechanism of the symmetric fitness variation, which was just assumed in the previous analyses. 
Our model demonstrates that such a variation is not an automatic consequence of efficient learning. 
Our result suggests that the symmetric variation can appear when Th clones have far more diversity than minimally required for achieving efficient learning under a given pathogen diversity (Fig. S4).
If the pathogen diversity is far beyond the capacity of the Th diversity, a part of the Th clones dominates in fitness than the others, which results in asymmetric fitness variation  (Fig. S4). 
While this problem is still open both theoretically and experimentally, our learning framework may provide a perspective from the viewpoint of computational and algorithmic levels.

\section*{Summary and Discussion}

The learning dynamics of the adaptive immune system has not yet been fully understood due to the lack of approaches at the computational and algorithmic levels despite the accumulated evidence at the implementation level. 
Based on the framework of reinforcement learning, we constructed a mathematical model, which may bridge this gap.
From our model, the clonal selection of Th cells is naturally derived as a learning rule, which enables the system not only to recognize new pathogens but also to acquire the appropriate way to bias the responses to the pathogens. 
Even though the simulations were conducted under an abstract and simple situation, we found a good scaling property among $K$, $N$, and $P$, which enables us to extrapolate our results to a more realistic scale.
In addition, our model could successfully reproduce the experimental clone size distributions to a certain extent when a sufficiently diverse Th clone type was assumed. 

Besides these results, our model still has room for accommodating more detailed quantitative information, if provided in the future, to make the simulation more realistic for more specific purposes.
For example, the dynamics of antigen patterns can be more detailed for describing specific infectious and pathological situations.
\TJKK{We may represent chronic infections by introducing hidden states in the dynamics of antigen patterns.
For such cases, more predictive behaviors with $\gamma>0$ might be related to actual immunological learning.
We can also introduce pre-training to mimic and analyze the thymic selection\cite{Klein2014NatRevImmunol}.}
Furthermore, if comprehensive quantitative data on the interactions between antigens and Th clones are obtained by future measurement technologies\cite{Newell2014NatBiotech,Kisielow2019NatImmunol,Jurtz2018bioRxiv,Glanville2017Nature}, we can include that information on the weights of the network. 

Nonetheless, we acknowledge that there are several discrepancies between the actual immune system and our mathematical model. 
First, the dynamics of the pathogen is implicit in our model because the framework of the MDP requires the agent (Th cell population) to be accessible to the environmental state\cite{Puterman2014Book,sutton1998reinforcement,hausknecht2015deep}. 
Such a problem can be addressed by extending the model to the partially observable MDP \cite{Puterman2014Book}. 
Second, the Th clones should be classified explicitly by the TCR and phenotypic state to directly compare the simulation to the experimental data.
Third, while the derived learning dynamics was qualitatively consistent with the clonal selection theory, the local feedback interactions from the effector cells to the Th cells should be associated with actual cell types and interacting molecules\cite{voisinne2015t, tkach2014t}. Similarly, the biological counterpart of the global signal and reward should be identified. Because the system cannot learn without the global reward signal, its identification can be a pivotal target for the verification of theoretical prediction.
In addition, the effector cells have an innate ability to recognize and respond to pathogens. 
Such effect is abstractly represented by the stochastic activation and inactivation of the effector cells in our model and is important for preventing the learning from becoming stuck in a certain infectious state. We may improve our model to involve more detailed and active behaviors of the effector cells as self-supporting agents. 
Such a hierarchical architecture resembles the memetic algorithm used for optimization \cite{Moscato1989}, and its investigation may deepen our understanding of the interrelationship between innate and adaptive immunity. 

Finally, while our model can share the characteristic feature of experimentally observed clone size distributions, it is not yet clear how the feature is related to the  general property of learning dynamics. 
Revealing the general aspects of abstract learning systems is also essential for understanding both universal and problem-specific properties of the immune system.

\begin{acknowledgments}
We greatly thank Yuki Sughiyama and Ryo Yokota for fruitful discussions. This research is supported by JSPS KAKENHI Grant Number 18H04814 and 18H05096, and by JST PRESTO Grant Number JPMJPR15E4, Japan.
\end{acknowledgments}

\appendix

\section{Simulation}

The simulation starts with a set of initializations. 
$P$ different realizable antigen patterns are initialized by random selections of $N$-dimensional binary vectors. 
Additionally, $P$ most effective activity patterns of the effector cells are initialized by randomly selecting $M$-dimensional binary vectors. 
The weights of the TCR-antigen interactions $\mathbf{w}$ are generated randomly by sampling each element of the matrix from a normal distribution $\mathcal{N}(0, \sigma^{2}_{w})$, where the variance $\sigma^{2}_{w}=2/N$ is determined by the He normal initialization method that is widely used in the context of deep learning\cite{he2015delving}. 
Each element in the weights $\mathbf{u}$ representing the strength of the signals from the Th clones to the effector cells is also sampled from a normal distribution $\mathcal{N}(0, \sigma^{2}_{u})$, where $\sigma^{2}_{u}=2/K$. 
The initial population size of each clone is uniformly set to $1$. 
The initial antigen pattern $\mathbf{s}(t)$ at $t=0$ is chosen uniformly at random.

The simulation was conducted by iterating the following steps. 
At each time step $t$, the activation pattern of the effector cells $\mathbf{a}(t)$ was determined by sampling from the policy $\pi_{\mathbf{n}(t)}(\mathbf{a}(t) | \mathbf{s}(t))$ calculated based on the antigen pattern $\mathbf{s}(t)$. 
The calculation of the policy depends on the global scaling parameter $\beta(t)$, which was gradually increased from $1.0$ to $20.0$ linearly towards the end of the iterations. 
Based on the activation pattern of the effector cells $\mathbf{a}(t)$, the reward $r(t)$ was determined as shown in \eqref{eq:reward}. 
The population size of each clone changes according to \eqref{eq:learn_eq} with a learning rate of $\alpha = 0.1$. 
If a clone size becomes lower than $0$, which is possible due to the time discretization in the simulation, the clone size is set to be $0$. 
Finally, the subsequent iterations were started after sampling the next antigen pattern $\mathbf{s}(t+1)$ from the transition probability \eqref{eq:transition_prob}.

All simulations were implemented either in MATLAB (R2018b; The MathWorks, Natick, MA) or Python using the standard scientific libraries numpy and scipy.

%

\end{document}